\documentclass[hyper]{JHEP3}
\pdfoutput=1
\usepackage{amsmath,amssymb,epsfig}
\usepackage{amsfonts}
\usepackage{verbatim}
\usepackage{latexsym}
\usepackage{amsthm}
\usepackage{amsbsy}
\usepackage{multirow}

\def\one{{\hbox{ 1\kern-.8mm l}}}
\newcommand{\Dslash}{\not{\hbox{\kern-4pt $D$}}}
\newcommand{\cL}{\mathcal{L}}

\newcommand{\Tr}{\mathrm{Tr}}

\newcommand{\SO}{\mathrm{SO}} 
\newcommand{\SU}{\mathrm{SU}} \newcommand{\U}{\mathrm{U}}

 \newcommand{\ie}{\emph{i.e.}\:}
 \newcommand{\pd}{\partial}

\newcommand{\Comment}[1]{{}}

\def\IZ{{\mathbb Z}}

\def\IR{{\mathbb R}}

\newcommand{\ra}{\rightarrow}

\newcommand{\bc}{\begin{center}}
\newcommand{\ec}{\end{center}}
\newcommand{\ba}{\begin{array}}
\newcommand{\ea}{\end{array}}
\newcommand{\beq}{\begin{equation}}
\newcommand{\eeq}{\end{equation}}
\newcommand{\bea}{\begin{eqnarray}}
\newcommand{\eea}{\end{eqnarray}}
\newcommand{\bmx}{\begin{pmatrix}}
\newcommand{\emx}{\end{pmatrix}}

\newcommand{\ep}{\epsilon}
\newcommand{\bep}{{\overline\epsilon}}
\newcommand{\vep}{\varepsilon}

\newcommand{\del}{\partial}
\newcommand{\half}{\frac{1}{2}}

\newcommand{\bPsi}{{\overline \Psi}}

\newcommand{\eref}[1]{Eq.~(\ref{#1})}

\newcommand{\tA}{{\tilde A}}

\def\IB{\relax{\rm I\kern-.18em B}}
\def\IC{{\relax\hbox{\kern.3em{\cmss I}$\kern-.4em{\rm C}$}}}
\def\ID{\relax{\rm I\kern-.18em D}}
\def\IE{\relax{\rm I\kern-.18em E}}
\def\IF{\relax{\rm I\kern-.18em F}}
\def\II{\relax{\rm I\kern-.18em I}}
\def\IZ{\relax{\sf Z\kern-.35em Z}}
\def\Id{\relax{1\kern-.32em 1}}
\def\IG{\relax\hbox{$\inbar\kern-.3em{\rm G}$}}
\def\IR{\relax{\rm I\kern-.18em R}}

\normalsize

\title{M2 to D2} 

\author
{Sunil Mukhi\,\footnote{Email: mukhi@tifr.res.in}\, and Constantinos
Papageorgakis\,
\footnote{Email: costis@theory.tifr.res.in}\\ \it Tata
  Institute of Fundamental Research,\\ \it Homi Bhabha Rd, Mumbai 400
  005, India}

\abstract{We examine the recently proposed ``3-algebra'' 
field theory for multiple M2-branes  and show that
when a scalar field valued in the 3-algebra
develops a vacuum expectation value, the resulting Higgs mechanism has
the novel effect of promoting topological (Chern-Simons) to dynamical
(Yang-Mills) gauge fields. This leads to a precise derivation of the
maximally supersymmetric Yang-Mills theory on multiple D2-branes and 
thereby provides a relationship between 3-algebras and Yang-Mills
theories. We discuss the physical interpretation of this result.}

\preprint{TIFR/TH/08-08}


\begin{document}

\section{Introduction}
\label{Introduction}

The world-volume theory on multiple M2-branes has remained mysterious
since the inception of M-theory over a decade ago. It is expected to
be the conformal-invariant IR fixed point of the D2-brane world-volume
theory, which to lowest order is a maximally supersymmetric Yang-Mills
theory in $2+1$ dimensions. The M2 theory should have 8 transverse
scalar fields as its bosonic content, while the D2 theory is known to
have 7 scalar fields and a gauge field (for a pedagogical review of
M-theory see\;\cite{Becker:2007zj}, for a review of M-branes see
Ref.\;\cite{Berman:2007bv}).

For the Abelian case, both the D2 theory and the M2 theory are free
and in this case the relation between them follows by performing an
Abelian duality on the gauge field of the D2 brane, which converts it
into the 8th scalar on the
M2-brane\;\cite{Duff:1992hu,Douglas:1995bn,Townsend:1995af,Schmidhuber:1996fy}. The
analogue of this relation has not been found for the non-Abelian case
so far, given the absence of a known interacting CFT for multiple
M2-branes.

Recently a concrete proposal has been made for the world-volume theory
on multiple M2 branes\;\cite{Bagger:2007jr,Gustavsson:2007vu} following
preliminary ideas in Refs.\;\cite{Basu:2004ed,Bagger:2006sk}. In this
proposal the field content is a collection of scalars, fermions and
gauge fields transforming under a ``3-algebra'', a generalisation of a
Lie algebra with a triple bracket replacing the commutator and a
4-index structure constant replacing the usual 3-index structure
constant of a Lie algebra. There is also a bilinear ``fundamental
identity'' replacing the Jacobi identity of a Lie algebra.\footnote{In
our work we use the notation, conventions and terminology of
Ref.\;\cite{Bagger:2007jr}.} The scalars and fermions are dynamical and
coupled via a sextic self-coupling and a 2-scalar-2-fermion analogue
of a Yukawa coupling. The gauge field, in contrast, is topological and
has a Chern-Simons self-coupling as well as minimal couplings to the
matter fields. It contributes no on-shell degrees of freedom.

The proposed action is maximally supersymmetric (the supersymmetry
algebra closes on-shell) and classically conformal invariant. It has
no free parameters and the structure constants of the 3-algebra are
quantised\;\cite{Bagger:2007vi}, strongly suggesting that conformal
invariance is exact at the quantum level. The theory has an elegant
and unique structure which makes it a very compelling candidate to be
the multiple M2-brane theory. In addition it has some features which
might not have been anticipated on general grounds, for example the
gauge symmetry associated to the Chern-Simons gauge field.

Nevertheless the proposal is incomplete for a few reasons. Only a
single 3-algebra (called ${\cal A}_4$) is explicitly known, and the
vacuum moduli space of the postulated theory has two free
parameters. If we add a zero mode supermultiplet (having vanishing
3-algebra bracket with all other fields) following the usual procedure
in D-brane theories, we get altogether three parameters, which has
been interpreted in Ref.\;\cite{Bagger:2007vi} as corresponding to
{\it three} M2-branes. A possible interpretation suggested there was
that there is no interacting theory for two M2-branes, and therefore
the IR limit of maximal SYM in $2+1$ dimensions is trivial, a
rather dramatic hypothesis for which no evidence is known.\footnote{We
are grateful to Shiraz Minwalla for stressing this point.} Another,
apparently independent, limitation of the proposed M2-brane theory is
that despite some
attempts\;\cite{Gustavsson:2007vu,Gustavsson:2008dy}, it has not been
possible to recover the multiple-D2-brane theory from it after
compactifying a transverse direction.

In this work we make an observation that relates the proposed multiple
M2-brane theory to the strongly coupled maximally supersymmetric
Yang-Mills theory in 2+1 dimensions.
This happens when a scalar field {\it develops a VEV in a 3-algebra
direction}. The resultant Higgsing leads to an $\SU(2)$ D2-brane
theory including a {\it dynamical} $\SU(2)$ gauge field, plus a
decoupled Abelian degree of freedom. Pleasingly, the $\SU(2)$ gauge
field is a part of the Chern-Simons gauge field of the original
3-algebra theory, which becomes dynamical via a generalised Higgs
mechanism as we will demonstrate. All interactions of the Yang-Mills
theory, and no others, are found in the strong coupling
limit. Finally, with minor changes the proposal extends to other
3-algebras, whose properties we can characterise but of which explicit
examples are not yet known.

In what follows we review the multiple M2-brane action based on
3-algebras, then describe our results in some detail for the 3-algebra
${\cal A}_4$. Next we discuss the generalisation to arbitrary
3-algebras and conclude with some open questions.

\section{The 3-algebra field theory}

The maximally supersymmetric 3-algebra field theory in $2+1$
dimensions\;\cite{Bagger:2007jr,Gustavsson:2007vu,Bagger:2007vi} describes a
set of bosonic fields $X^{A(I)}, A_\mu^{~AB}$ and fermionic fields
$\Psi^A$ having (suppressed) spinor indices with respect to $\SO
(2,1)$ as well as $\SO (8)$. Here the indices $\{A,B,...\}$ take the
values $1,...,\rm{dim}_{\mathcal A}$ with $\mathrm{dim}_{\mathcal A}$
being the dimension of a 3-algebra, which we will leave unspecified
for the moment, while $\{I,J,...\}=1,2,\ldots , 8$ label the scalar
fields corresponding to the 8 directions transverse to the M2-branes.

To write the action we first introduce 4-index structure constants
$f^{ABCD}$ associated with a formal, totally antisymmetric
three-bracket over the three-algebra generators:
\begin{equation}
[T^A, T^B, T^C] = f^{ABC}_{\phantom{ABC}D}T^D
\end{equation}
and a generalisation of the trace, ``Tr'' taken over the three-algebra
indices, which provides an appropriate `3-algebra metric':
\begin{equation}
 h^{AB} = \Tr(T^A, T^B)\;.
\end{equation}
Then the 4-index structure constants  satisfy the `fundamental identity':
\begin{equation}
\label{funid}
f^{AEF}_{~~~~~G}\,f^{BCDG}-
f^{BEF}_{~~~~~G}\,f^{ACDG}+
f^{CEF}_{~~~~~G}\,f^{ABDG}-
f^{DEF}_{~~~~~G}\,f^{ABCG}=0
\end{equation}
and are also completely antisymmetric under the exchange of indices:
\beq
f^{ABCD} = f^{[ABCD]}\;.
\eeq

All information about the 3-algebra is contained in the structure
constants, so we will write actions and equations of motion without
referring again to the 3-bracket. This avoids the question of what
algebraic structure (analogous to matrices) is encoded in the
3-bracket, and removes much of the mystery from the action -- which is
ultimately a set of couplings among multiplets of ordinary fields. This
action is:
\begin{equation}
\begin{split}
{\cal L} &= -\half D_\mu X^{A(I)}D^\mu X_A^{(I)} +
\frac{i}{2}{\bPsi}^A\Gamma^\mu D_\mu \Psi_A
+\frac{i}{4} f_{ABCD} \bPsi^B \Gamma^{IJ}X^{C(I)}X^{D(J)}\Psi^A \\
&~~~ -\frac{1}{12} \left(f_{ABCD}X^{A(I)}X^{B(J)} X^{C(K)}\right)
\left(f_{EFG}^{\phantom{EFG}D}X^{E(I)}X^{F(J)}X^{G(K)} \right)\\
&~~~ +\half\,\vep^{\mu\nu\lambda}\left(
f_{ABCD}A_\mu^{~AB}\del_\nu A_{\lambda}^{~CD}
+ \frac23 f_{AEF}^{~~~~~G}\,f_{BCDG}\,
A_{\mu}^{~AB}A_{\nu}^{~CD}A_{\lambda}^{~EF}\right)\\
\end{split}
\end{equation}
where:
\begin{equation}
D_\mu X^{A(I)} = \del_\mu X^{A(I)} + f^{A}_{~~BCD}A_\mu^{CD}X^{B(I)}\;.
\end{equation}
It is invariant under the gauge transformations:
\begin{equation}
\begin{split}
\delta X^{A(I)} &= -f^A_{~~BCD}\Lambda^{BC}X^{D(I)}\\
\delta \Psi^A & = -f^A_{~~BCD}\Lambda^{BC}\Psi^D\\
\delta (f_{AB}^{~~~\,CD}A_\mu^{AB}) &= f_{AB}^{~~~\,CD}D_\mu \Lambda^{AB} 
\end{split}
\end{equation}
and the supersymmetries:
\begin{equation}
\label{susies}
\begin{split}
\delta X^{A(I)} &= i\,\bep\,\Gamma^I\Psi^A\\
\delta \Psi^A & = D_\mu X^{A(I)}\Gamma^\mu \Gamma^I \ep +\frac16
f^A_{~~BCD}X^{B(I)}X^{C(J)}X^{D(K)} \Gamma^{IJK}\ep\\
\delta (f_{AB}^{~~~\,CD}A_\mu^{AB}) &= i f_{AB}^{~~~\,CD}
X^{A(I)}\,\bep\, \Gamma_\mu\Gamma_I
\Psi^B\\ 
\end{split}
\end{equation}
where $\Gamma_{012}\ep = \ep$ and $\Gamma_{012}\Psi^A = -\Psi^A$. 

A potentially puzzling feature of this theory is that while the
fundamental gauge field is $A_\mu^{AB}$, it is the combination
$\tA^{CD}_\mu=f_{AB}^{~~~\,CD}A_\mu^{AB}$ that appears in the symmetry
transformations and covariant derivatives, despite the fact that the
Chern-Simons action cannot be written in terms of $\tA$ alone. This
was explained in Ref.\;\cite{Bagger:2007jr} by noting that those
variations in $A$ that do not affect $\tA$ leave the Chern-Simons
action invariant. Therefore in a subtle way, the theory depends only
on the gauge field $\tA$.

The above theory is manifestly conformally invariant at the
classical level. If the proposal that it describes M2-branes is
correct then it must also be quantum mechanically conformal
invariant. This is very plausible, though it has not yet been
explicitly demonstrated.

The one 3-algebra that can be easily constructed (in fact, the only
one constructed so far) has structure constants given by the 4-index
totally antisymmetric symbol $f^{ABCD}=\vep^{ABCD}$ with $A,B,C,D\in
\{1,2,3,4\}$. This is the lowest dimensional 3-algebra that one
can write down and has been denoted ${\cal A}_4$. The action in this
case has an $\SO(4)$ rotation invariance. For this action it was
observed in Ref.\;\cite{Bagger:2007vi} that the vacuum moduli space,
defined as the space of solutions to the equations:
\beq
f_{ABCD} X^{A(I)}X^{B(J)}X^{C(K)}=0
\eeq
is given by:
\beq
X^{A(I)} = a^{(I)} \alpha^A + b^{(I)} \beta^A
\eeq
where $\alpha^A, \beta^A$ are arbitrary elements of the ${\cal A}_4$
algebra and $a^I,b^I$ are constant vectors. It was postulated that
3-algebras for M2-branes should be supplemented by a new ``central''
direction ``0'' such that $f^{0ABC}=0$ for all $A,B,C$, and that the
fields $X^{0(I)},\Psi^0$ describe the overall centre-of-mass or zero
mode of the M2-brane system. Adding in the zero mode for the special
case of the ${\cal A}_4$ algebra, one finds a 3-parameter vacuum
moduli space that was interpreted in Ref.\;\cite{Bagger:2007vi} as
describing {\it three} M2-branes. We will re-examine this
interpretation in the concluding section.

\section{M2 to D2 for $\U(2)$}

In this section we re-examine the field theory based on the ${\cal
A}_4$ 3-algebra and will find that it quite naturally describes a pair
of 2-branes coupled via a supersymmetric Yang-Mills action, along
with a free Abelian theory.
The emergence of dynamical Yang-Mills interactions constitutes a
sensitive check of the proposed M2-brane action and tests many of its
detailed features, including its somewhat baroque Chern-Simons
structure.

We start by assuming a scalar field in the 3-algebra theory develops a
VEV equal to a length parameter $R$. Because of $\SO(4)$ invariance it
is possible to rotate the scalar field that gets a VEV to have only
the component $X^{4(8)}$.  In order to make the notation suitable for
the more general case, at this point we re-label the 3-algebra
direction ``4'' as ``$\phi$''. Thus the four indices split into
$a\in\{1,2,3\}$ and $\phi$. The direction $\phi$ singled out in this
manner will shortly be interpreted as the zero-mode.

Because scalar fields have canonical dimension $\half$ while $R$ has
dimension $-1$, our proposal amounts to saying that:
\begin{equation}
\langle X^{\phi(8)}\rangle = \frac{R}{\ell_p^{3/2}}\;.
\end{equation}
When compactifying M-theory on a circle of radius $R$ to type IIA
string theory, the RHS of the above equation turns out to equal
$\sqrt{\frac{g_s}{\ell_s}}\equiv g_{YM}$ where $g_s,\ell_s$ are the
string coupling and string length and $g_{YM}$ is the dimensional
coupling on D2-branes.

Let us now examine the theory with this VEV. To begin with, note that
a VEV $\langle X^{\phi(8)}\rangle$ preserves supersymmetry as long as
no other field has a VEV. To see this, consider the fermion variation
in \eref{susies}. The first term on the RHS is zero because the scalar
VEV is constant while the gauge field VEV is of course zero. The
second term vanishes because $X^{\phi(8)}$ can occur at most once in
it, while the other two scalar fields have a vanishing VEV. Therefore
the theory expanded about this scalar VEV will have maximal
supersymmetry. It also depends on a dimensional coupling constant
$g_{YM}$ of canonical dimension $\half$ as expected, and in agreement
with the fact that this theory is weakly coupled in the UV and
strongly coupled in the IR.

Now let us examine the various terms in the Lagrangean and show how
they reproduce the SYM theory. To start with, consider the sextic
potential. Introduce the labels $a,b,c\in \{1,2,3\}$ as well as
$i,j,k\in \{1,2,...,7\}$. Then the potential is:
\begin{equation}
\begin{split}
V(X) &=\frac{1}{12} \sum_{I,J,K=1}^8
\left(\vep_{ABCD} \vep_{EFG}^{\phantom{EFG}D}X^{A(I)}X^{B(J)}X^{C(K)}
  X^{E(I)}X^{F(J)} X^{G(K)}\right)\\
&=
\frac{1}{2} \sum_{i<j}^7
\left(\vep_{ABCD} \vep_{EFG}^{\phantom{EFG}D}X^{A(i)}X^{B(j)}X^{C(8)}
  X^{E(i)}X^{F(j)} X^{G(8)}\right)\\
&~~~+
\frac{1}{2} \sum_{i<j<k}^7
\left(\vep_{ABCD} \vep_{EFG}^{\phantom{EFG}D}X^{A(i)}X^{B(j)}X^{C(k)}
  X^{E(i)}X^{F(j)} X^{G(k)}\right)
\\
&= \frac{1}{2}\, g_{YM}^2 \sum_{i<j}^7
\left(\vep_{ab4d}\vep_{ef\phi}^{\phantom{ef\phi}d} 
X^{a(i)}X^{b(j)}X^{e(i)}X^{f(j)}\right)
+ g_{YM}{\cal O}\left(X^5\right) + {\cal O}\left(X^6\right)\;.
\end{split}
\end{equation}
In the last line we have inserted the VEV $\langle
X^{\phi(8)}\rangle=g_{YM}$, which leads to a term quartic in the
remaining $X$'s. Note that in this term, only $X^{a(i)}$ appear where
$a\in \{1,2,3\}$ and $i\in \{1,2,...,7\}$. The terms of order $g_{YM}
\mathcal O (X^5)$ and $\mathcal O(X^6)$ have not been written
explicitly because they will decouple at strong coupling.

Using $\vep_{abd\phi}\equiv\vep_{abd}$ where the latter is the 3-index
totally antisymmetric symbol and structure constant of an $\SU(2)$ Lie
algebra, we see that the quartic term becomes:
\begin{equation}
\frac{1}{2}\, g_{YM}^2 \sum_{i<j=1}^7
\left(\vep_{abc}\vep_{ef}^{\phantom{ef}c} 
X^{a(i)}X^{b(j)}X^{e(i)}X^{f(j)}\right)\;,
\end{equation}
which is precisely the quartic scalar interaction of maximally
supersymmetric $\SU(2)$ SYM in $2+1$ dimensions.

Following the same procedure, it is easy to check that the 2-fermion,
2-scalar coupling reduces to the Yukawa coupling of $2+1$
dimensional SYM, plus terms with two fermions and two scalars:
\begin{equation}
\frac{i}{4} \vep_{ABCD} \bPsi^B \Gamma^{IJ}X^{C(I)}X^{D(J)}\Psi^A 
= \frac{i}{2} g_{YM}\,\vep_{abc} \bPsi^b \Gamma^{i}X^{c(i)}\Psi^a
+ {\cal O}\left(X^2\Psi^2\right)\;.
\end{equation}
We see that the only scalars and fermions appearing in the first term
(which will be the leading term in the strong coupling limit) are
$\Psi^a$ and $X^{a(i)}$.

Since kinetic terms are unaffected by a scalar VEV, it only remains to
understand the gauge field terms including couplings of gauge fields
through covariant derivatives. On the face of it this should be the
major stumbling block, for the gauge field in the 3-algebra theory
only has Chern-Simons couplings while the D2-brane Yang-Mills theory
requires a dynamical gauge field.\footnote{Some kind of non-Abelian
duality like $D_\mu X^{a(8)}\sim
\vep_\mu^{~\nu\lambda}F^a_{\nu\lambda}$ has been proposed in the
past\;\cite{Gustavsson:2007vu} but so far this has not been possible
to implement precisely.} We will make no additional assumptions to
account for the dynamical gauge field, but simply work out the full
content of the theory in the presence of the VEV of the scalar field
$X^{\phi(8)}$. We will find that the Higgs mechanism, and the original
Chern-Simons coupling, miraculously conspire to provide the desired
dynamical gauge field with all the right properties.

In view of our split of indices $A,B\in \{1,2,3,4\}$ into $a,b\in
\{1,2,3\}$ and $\phi=4$, it is natural to break up the gauge field
$A_\mu^{AB}$ into two parts:
\begin{equation}
\begin{split}
A_\mu^{~a\phi}&\equiv A_\mu^{~a}\\
\half\vep^a_{~bc} A_\mu^{~bc}&\equiv B_\mu^{~a}\;.\\
\end{split}
\end{equation}
Each of these is a triplet of vector fields. We can now re-write the
two terms in the Chern-Simons action as follows:
\begin{equation}
\begin{split}
\half\,\vep^{\mu\nu\lambda}
\vep_{ABCD}A_\mu^{~AB}\del_\nu A_{\lambda}^{~CD}
&= 2\,\vep^{\mu\nu\lambda}\vep_{abc} A_\mu^{~ab}\del_\nu
A_\lambda^{~c}=
4\,\vep^{\mu\nu\lambda}\,
B_\mu^{~a} \del_\nu A_{\lambda\,a}\\
\frac13\,\vep^{\mu\nu\lambda}\,\vep_{AEF}^{~~~~~G}\,\vep_{BCDG}\,
A_\mu^{~AB}A_\nu^{~CD}A_\lambda^{~EF} &=
-4\, \vep^{\mu\nu\lambda}\,\vep_{abc} B_\mu^{~a}A_\nu^{~b}A_\lambda^{~c}
-\frac43\,\vep^{\mu\nu\lambda}\,\vep_{abc} 
B_\mu^{~a}B_\nu^{~b}B_\lambda^{~c}\;.
\end{split}
\end{equation}
We also need to consider the couplings arising from the covariant
derivative on $X^{A(I)}$. We have:
\begin{equation}
\label{covdev}
\begin{split}
D_\mu X^{a(I)}&=
\del_\mu X^{a(I)} + \vep^a_{~BCD}A_\mu^{~CD}X^{B(I)}\\
&= \del_\mu X^{a(I)} + 2\,\vep^a_{~bc}A_\mu^{~c}X^{b(I)}
+2\,B_\mu^{~a}X^{\phi(I)}
\end{split}
\end{equation}
and: \beq\label{covdev4} D_\mu X^{\phi(I)} = \del_\mu X^{\phi(I)}-2
B_{\mu a}X^{a(I)}\;.  \eeq Inserting these in the Lagrangean (but
ignoring fermions) and using the VEV $\langle
X^{\phi(8)}\rangle = g_{YM}$, we find the following terms involving
$B_\mu^{~a}$\;:
\begin{equation}
\begin{split}
  \mathcal L_{\rm kinetic} = -2 g_{YM}^2 B_\mu^{~a}B^\mu_a -2
  &B^{~a}_\mu X^{\phi(I)}D'^\mu X_a^{(I)}-2 g_{YM}B^{~a}_\mu
  D'^\mu X_a^{(8)}\\-2B_{\mu
    a}X^{a(I)}B^\mu_bX^{b(I)}
  &-2 B^a_{\mu }B^\mu_aX^{\phi(I)}X_{\phi(I)}+ 2 B^\mu_a X^{a(I)}
  \pd_\mu X^{\phi(I)} +... \;,
\end{split}
\end{equation}
where we have defined a new covariant derivative which depends only on
$A_\mu^a$\;:
\beq
D'_\mu X^{a(I)} = \pd_\mu X^{a(I)}-2 \vep^a_{\phantom{a}bc}A^b_\mu X^{c(I)}\;. 
\eeq
Notice that the first term is a mass for $B_\mu^{~a}$, as one would
expect from the Higgs mechanism.

Similarly, the terms involving $B_\mu^a$ which come from  the gauge field
self-couplings are:
\begin{equation}
\cL_{\rm CS}  = 2\,\vep^{\mu\nu\lambda}\,
B_\mu^{~a} F'_{\nu\lambda a}
-\frac43\,\vep^{\mu\nu\lambda}\,\vep_{abc} 
B_\mu^{~a}B_\nu^{~b}B_\lambda^{~c} +...\;,
\end{equation}
where we have also defined: 
\beq F_{\nu\lambda}^{'a} = \pd_\nu
A_\lambda^a - \pd_\lambda A^a_\nu - 2\vep^a_{\phantom{a}b c}A^b_\nu
A^c_\lambda\;.
\eeq
 Notice that by virtue of its Chern-Simons nature,
$B_\mu^{~a}$ is an auxiliary field appearing without derivatives. It
can therefore be eliminated via its equation of motion. We can extract
the leading part of such solution by temporarily neglecting the
quadratic term in $B_\mu^a$ coming from the cubic self-interaction as
well as terms coming from higher interactions with
scalars. Later we will show that these  would have led to
higher-order contributions which are suppressed in the strong coupling
limit.  We therefore consider the set of couplings:
\beq
\cL =  -2 g_{YM}^2 B_\mu^{~a}B^\mu_a -2 g_{YM}B^{~a}_\mu
  D'^\mu X_a^{(8)} + 2\,\vep^{\mu\nu\lambda}\,
  B_\mu^{~a} F'_{\nu\lambda a} + \hbox{higher order}
\eeq
and find that:
\begin{equation}
B_\mu^{~a} =
\frac{1}{2g_{YM}^2}
\vep_\mu^{\phantom{\mu}\nu\lambda}\,F_{\nu\lambda}^{'a} 
- \frac{1}{2g_{YM}}D'_\mu X^{a(8)}\;.
\end{equation}

Thus one of our gauge fields, $B_\mu^{~a}$, has been set equal to the
field strength of the other gauge field $A_\mu^{~a}$ (plus other
terms). Together with the fact that $B_\mu^{~a}$ has a mass term, we
now see that eliminating $B_\mu^{~a}$ will provide a standard
Yang-Mills kinetic term for $A_\mu^{~a}$! This is the desired miracle
that promotes the Chern-Simons gauge field $A_\mu^{~a}$ into a
dynamical gauge field.

Continuing with the computation, the sum of the Chern-Simons gauge
field action and the scalar covariant kinetic terms becomes (up to a
total derivative):
\begin{equation}\label{kinetic-cs}
-\frac{1}{g_{YM}^2} F^{'a}_{\mu\nu} F^{'\mu\nu}_a
- \frac{1}{2}\pd_\mu X^{\phi(I)}\pd^\mu X_\phi^{(I)}
- \frac{1}{2}D_\mu X^{a(i)}D^\mu X_a^{(i)}
+\mathcal O (B X\pd X)+\mathcal O (B^2 X^2) + \mathcal O (B^3)\;.
\end{equation}
A re-definition:
\beq
A\ra \frac{1}{2}A\;,
\eeq
leads to:
\beq
 D'_\mu X^{a(I)}\ra D_\mu X^{a(I)} \equiv \pd_\mu X^{a(I)} -
 \vep^a_{~bc}A_\mu^b X^{c(I)}
\eeq
and:
\beq
 F'^a_{\mu\nu}\ra \frac{1}{2}F^a_{\mu\nu}\equiv
\frac{1}{2}\left(\pd_\mu A^{~a}_\nu - \pd_\nu A^{~a}_\mu -
 \vep^a_{~bc}A^{~b}_\mu A^{~c}_\nu\right) \;.
\eeq
Thus \eref{kinetic-cs} finally becomes:
\beq
\begin{split}
&-\frac{1}{4 g_{YM}^2} F^a_{\mu\nu} F^{\mu\nu}_a
- \frac{1}{2}\pd_\mu 
X^{\phi(I)}\pd^\mu X_\phi^{(I)}
- \frac{1}{2}D_\mu X^{a(i)}D^\mu X_a^{(i)}+\frac{1}{g_{YM}}\mathcal
O\left( X\pd X \left(F/g_{YM}+ DX\right)\right)\\
&~~~~~+\frac{1}{g^2_{YM}}\mathcal O \left(X^2\left(F/g_{YM}+ DX\right)^2\right)
 + \frac{1}{g_{YM}^3}\mathcal O
\left(\left(F/g_{YM}+ DX\right)^3\right)\;.\\
\end{split}
\eeq
The terms in $B^a_\mu$ that we had neglected will lead to higher
interactions with increasingly higher powers of $(F/g_{YM}+D X)$ in
the numerator and $g_{YM}$ in the denominator.

For the fermions, we easily find that:
\beq
\frac{i}{2}{\bar\Psi}^A \Gamma^\mu D_\mu \Psi_A \to
\frac{i}{2}{\bar\Psi}^a \Gamma^\mu D_\mu \Psi_a +
\frac{i}{2}{\bar\Psi}^\phi \Gamma^\mu \del_\mu \Psi_\phi
+ \hbox{higher order}\;,
\eeq 
where $D_\mu$ on the LHS is the 3-algebra covariant derivative while
$D_\mu$ on the right is the Yang-Mills covariant derivative.

The theory we have obtained now has conventional $\SU(2)$ Yang-Mills
couplings supplemented with some decoupled fields as well as a variety
of higher-order terms.\footnote{The original 3-algebra still makes its
presence in the higher-order terms.} The action can be written in the
form:
\beq
\cL=\cL_{\rm decoupled}+\cL_{\rm coupled}
\eeq
where
\beq
\cL_{\rm decoupled}= -\half\del_\mu X^{\phi(I)}\del^\mu X^{(I)}_{\phi}
+ \frac{i}{2}{\bar \Psi}^\phi \Gamma^\mu\del_\mu\Psi_\phi\;.
\eeq
For the interacting part, we re-scale the fields as $(X,\Psi)\to
(X/g_{YM}, \Psi/g_{YM})$, to find the action:
\beq
\label{lcoupled}
\cL_{\rm coupled} 
= \frac{1}{g_{YM}^2}\cL_{0} + \frac{1}{g_{YM}^3}\cL_{1} + {\cal
O}\left(\frac{1}{g_{YM}^4}\right) 
\eeq
where $\cL_0$ is the action of maximally supersymmetric $2+1$
dimensional Yang-Mills theory:
\begin{equation}\label{D2}
\begin{split}
\cL_0 = & -\frac{1}{4}F_{\mu\nu\,a}F^{\mu\nu\,a}-\frac{1}{2}D_\mu
X^{a(i)}D^\mu X_a^{~(i)} + \frac{1}{4}
\left(\vep_{abc}X^{a(i)}X^{b(j)}\right)
\left(\vep_{de}^{~~c}X^{d(i)}X^{e(j)}\right)\\
& +\frac{i}{2}\bar
\Psi^a \Dslash\Psi_a + \frac{i}{2}\vep_{abc}
{\bar\Psi}^a \Gamma^i X^{b(i)}\Psi^c\;,
\end{split}
\end{equation}
with the field strength and covariant derivative defined as:
\begin{equation}
F^a_{\mu\nu} = \partial_\mu A^a_\nu - \partial_\nu A^a_\mu -
\vep^a_{~bc}A^b_\mu A^c_\nu\quad \textrm{and}\quad D_\mu^{ab} = 
\partial_\mu\delta^{ab} + \vep^{ab}_{~~c}A^c_\mu\;.
\end{equation} 
Note that in the above, $\cL_0,\cL_1,...$ are all completely
independent of $g_{YM}$.

Since the M2-brane is supposed to describe the strongly coupled
Yang-Mills theory, we expect it will match on to SYM in the IR limit
$g_{YM}\to\infty$. In this limit, we see from \eref{lcoupled} that the
interacting part of the surviving theory is precisely the $\SU(2)$ SYM
theory on two D2-branes. Note that $B_\mu^{~a}$ has disappeared from
the theory while $A_\mu^{~a}$ no longer has a Chern-Simons coupling
but rather a full-fledged $\SU(2)$ Yang-Mills action. The fields that
survive in the D2-brane action have precisely the right
covariant-derivative couplings to the newly-dynamical gauge field. The
right quartic and Yukawa couplings have already been obtained at the
beginning of this section. Finally, the terms corresponding to the
modes $X^{a(8)}$ have disappeared; they have played the role of the
Goldstone bosons that gave a mass to $B^a_\mu$ and at the end, have
transmuted via the Higgs mechanism and the Chern-Simons coupling into
the single physical polarisation of $A_\mu^{~a}$.

Our final theory also contains 8 non-interacting scalars
$X^{\phi(I)}$.  Of these, $X^{\phi(i)}, i=1,2,...,7$ correspond to the
centre-of-mass modes for the D2 world-volume theory. The last scalar
$X^{\phi(8)}$, the one which originally developed a VEV, can now be
dualised via an {\it Abelian} duality to yield an extra $\U(1)$ gauge
field. The free Abelian multiplet is completed by $\Psi^\phi$. The
whole multiplet comes from a direction that was {\it not central} in
the original 3-algebra.

One might be alarmed at the fact that the original gauge symmetry
$\SO(4)\simeq \SU(2)\times \SU(2)$ appears to have been Higgsed to
$\SU(2)\times \U(1)$ by a VEV of a field in the $4$ of $\SO(4)$. That is
not quite the case. The Higgs mechanism breaks $\SO(4)$ to $\SO(3)\simeq
\SU(2)$ as it should, but several free scalars are left over, and the
$\U(1)$ gauge field is obtained by dualising one of them.

\section{M2 to D2: general case}

In this section we extend our proposal to more general 3-algebras. We
will be hampered by the scant knowledge of 3-algebras but will find
that the general case proceeds in much the same way as the
${\cal A}_4$ case that we just examined, though there are also some
differences.

We start by observing a key  feature of the `fundamental identity'
\eref{funid}.\footnote{We are grateful to Neil Lambert for emphasising
this to us.} By fixing two of the indices to take a specific value,
say $E=A=\phi$, and defining 3-index structure constants via
$f^{abc}\equiv f^{abc\phi}$, one recovers the Jacobi identity for the
usual Lie algebras:
\beq 
f^{dfg}f^{bc}_{\phantom{bc}g}+
f^{bfg}f^{cd}_{\phantom{bc}g}+
f^{cfg}f^{db}_{\phantom{bc}g}=0\;.
\eeq
The indices $\{a,b,...\} = 1,...,\rm{dim}_{\mathcal A}-1$ run over the
dimension of the Lie algebra. We will call this
\emph{$\mathcal Q$}, \ie $\rm{dim}_{\mathcal A}-1 
= \rm{dim}_{\it{\mathcal Q}}$.

In view of our preceding observations, we would like to interpret this
fact as saying that after assigning a VEV to one scalar,
the remaining directions describe $\SU(N)$ degrees of freedom coupled
via an SYM theory. This provides some constraints on the 3-algebra,
namely all the structure constants $f^{abc\phi}$ are
determined. However, as we now see, there are more 3-algebra structure
constants to be determined.

Recall that for the ${\cal A}_4$ 3-algebra, the structure constants
$\vep^{ABCD}$ reduced to $\vep^{abc\phi}$, with $a,b,c\in\{1,2,3\}$
and obviously there were no components $\vep^{abcd}$ left over. This
is less obvious in the general case, for which we can allow $f^{ABCD}$
to split into both $f^{abc\phi}$ as well as $f^{abcd}$. Indeed, if the
algebra is not ${\cal A}_4$ then $f^{abcd}$ cannot all be zero as we
now show. For this, assume the contrary, namely that $f^{abcd}=0$ for all
$a,b,c,d\in \{1,2,...,{\mathcal A}-1\}$. In that case choosing
$A,B,C,D,E,F$ to be $a,b,c,d,e,f$ in \eref{funid}, we find that the
summation index $g$ can only be equal to $\phi$. As a result we have
the identity:
\beq
f^{aef}\,f^{bcd}-
f^{bef}\,f^{acd}+
f^{cef}\,f^{abd}-
f^{def}\,f^{abc}=0\;.
\eeq 
This identity, involving no summation over common indices, certainly
does not hold for the structure constants of general Lie
algebras. However it does hold for $f^{abc}=\vep^{abc}$ just because
the number of possible indices is so small. This shows that in general
the assumption $f^{abcd}=0$ is incompatible with the fundamental
identity. Hence, in the general case we will have 3-algebra
interactions even among the $X^{a(I)}$s.

To recover a D2-brane gauge theory with gauge group $\SU(N)$, we assume
there exists a 3-algebra with structure constants
$f^{ABCD},~A,B,C,D\in\{1,2,..., N^2\}$. We next pick some
direction $\phi$ and identify the structure constants $f^{abc\phi}$
with the $f^{abc}$ of $\SU(N)$, where $a,b,c\in \{1,2,...,N^2-1\}$.
We can now, as before, break up the scalar fields $X^{A(I)}$ and the
fermions $\Psi^A$ into the sets $X^{a(I)},X^{\phi(I)}$ and
$\Psi^a,\Psi^\phi$. The first step of our reduction procedure is then
to postulate that:
\beq
\langle X^{\phi(8)}\rangle=g_{YM}\;.
\eeq
Expanding around this VEV, the sixth order interactions descend to
quartic plus higher-order terms:
\begin{equation}
\frac{1}{2}\, g_{YM}^2 \sum_{i<j=1}^7
\left(f_{abc}f_{ef}^{\phantom{ef}c} 
X^{a(i)}X^{b(j)}X^{e(i)}X^{f(j)}\right) + ...\;,
\end{equation}
just as in the ${\cal A}_4$ case. Reduction of the
two-fermion, two-scalar coupling to the Yukawa coupling proceeds in
the same manner.

Thus the only new feature arises with the gauge fields
$A_\mu^{AB}$. We can again split them into the two sets:
\begin{equation}
A_\mu^{~a\phi}\equiv A_\mu^{~a},\qquad A_\mu^{~bc}\;.\\
\end{equation}
Now  we naively no longer have equal numbers of components in the two
sets. The single-index field $A_\mu^{~a}$ has $N^2-1$ components.  The
other field $A_\mu^{~bc}$ has instead $\frac{(N^2-1)(N^2-2)}{2}$
components, which equals $N^2-1$ only for $N^2=4$ which is the $\SU(2)$
case. In general it has many more components than $A_\mu^{~a}$. This
appears to contradict the idea of making one part of the gauge field
massive via the Higgs mechanism and then, by eliminating that field,
rendering the other one dynamical.

However, we are saved by a property of the theory referred to earlier.
When we consider the covariant derivatives, we see that the only
combinations of gauge fields that appear are $A_\mu^{~a}$ and 
$B_\mu^{~a} = \half f^a_{~bc}A_\mu^{~bc}$:
\beq
D_\mu X^{a(I)} = \del_\mu X^{a(I)} + f^a_{~BCD} A_\mu^{CD} X^{B(I)}
= \del_\mu X^{a(I)} + 2f^a_{~bc}A_\mu^{~c} X^{b(I)}
+ 2B_\mu^{~a}X^{\phi(I)}\;.
\eeq
This is a manifestation of the fact\;\cite{Bagger:2007jr} that the
theory depends only on $\tA$ rather than $A$.

Similarly when we examine the Chern-Simons couplings, we find that the
combination of $A_\mu^{ab}$ that couples to $A_\mu^a$ is precisely:
\beq
B_\mu^{~a}\del_\nu A_{\lambda ~a}
\eeq
plus cubic terms of the form $B\wedge A\wedge A$. Therefore our
previous procedure goes through essentially unchanged.  The Higgs
mechanism causes $X^{a(8)}$ to disappear from the spectrum by giving a
mass to $B_\mu^{~a}$, and this field is then eliminated by setting it
equal to the field strength $F_{\mu\nu}^{~a}$, leading to the
promotion of $A_\mu^{~a}$ to a dynamical gauge field. A set of free
fields $X^{\phi(I)}$ and $\Psi^\phi$ are left over to generate the
decoupled $\U(1)$ multiplet. The resulting theory
therefore has an $\SU(N)$ Yang-Mills part that survives at strong
coupling, and an Abelian part.

The above procedure is less explicit only to the extent that a
construction of the 3-algebra structure constants is not
known. However it suggests a way to proceed. Given that
$f^{abc\phi}$ are completely known, we consider them as ``input'' for
the set of linear equations obtained by putting {\it one} free index
in the fundamental identity equal to $\phi$:
\begin{equation}
f^{aef}_{~~~\,g}\,f^{bc\phi g}-
f^{bef}_{~~~\,g}\,f^{ac\phi g}+
f^{cef}_{~~~\,g}\,f^{ab\phi g}-
f^{\phi ef}_{~~~\,g}\,f^{abcg}=0
\end{equation}
which can be re-written as:
\begin{equation}
f^{bcg}\, f^{aef}_{~~~\,g}+
f^{cag}\, f^{bef}_{~~~\,g}+
f^{abg}\, f^{cef}_{~~~\,g}=
f^{efg}\, f^{abc}_{~~~\,g}\;.
\end{equation}
Treating the 3-index $f^{abc}$ as input, this is a set of linear
equations for the unknown quantities $f^{abcd}$. Solutions to this
system of equations should be easier to classify, because of
linearity, than solutions of the full fundamental identity. In this
sense, the reduction to $\SU(N)$ structure constants when one index is
set equal to $\phi$ is like a boundary condition. Finally, one has to
ensure that the resulting structure constants satisfy the full
fundamental identity, which for the reduced $f^{abcd}$ becomes:
\begin{equation}
f^{aef}_{~~~\,g}\,f^{bcdg}-
f^{bef}_{~~~\,g}\,f^{acdg}+
f^{cef}_{~~~\,g}\,f^{abdg}-
f^{def}_{~~~\,g}\,f^{abcg}= -(f^{aef}\,f^{bcd}-
f^{bef}\,f^{acd}+
f^{cef}\,f^{abd}-
f^{def}\,f^{abc})
\end{equation}
We hope to carry out this analysis in the future. 

\section{Discussion}

We have shown that when one component $X^\phi(8)$ of the scalar fields
develops a VEV proportional to $R$, the ensuing Higgs mechanism
produces a strongly coupled SYM theory on a {\it pair} of D2-branes,
along with a decoupled theory. The emergence of SYM, complete in all
details, from the 3-algebra affirms the relationship of 3-algebra
theories to string theory and thereby M-theory. Every interaction of
the 3-algebra theory is tested, including its most unusual feature of
a Chern-Simons field with a gauge group under which the physical
fields apparently transform as fundamental rather than adjoint
fields. After Higgsing, part of the Chern-Simons gauge field has
become dynamical, and the physical fields are adjoints of this
dynamical gauge field.

Let us discuss the physical interpretation of our results.\footnote{We
are grateful to Shiraz Minwalla, Ashoke Sen and David Tong for their
helpful comments on a first version of this manuscript.} The
emergence of D2-brane theories may suggest we are dealing with a
compactification of M-theory. Upon compactifying M-theory on a circle
of radius $R$, there will be a periodic array of M2-branes in the
$x^8$ direction. When dealing with D-branes in string theory, one can
derive the dynamics explicitly following Ref.\;\cite{Taylor:1996ik}. An
infinite periodic array of the D-brane system along the chosen
direction causes the finite matrices on the world-volume to be
extended to $\infty\times
\infty$ matrices that incorporate the degrees of freedom of strings
connecting branes at different places in the periodic array. The
result is then an $R$-dependent action. Next, one quotients both the
space and the world-volume theory by a translation, which compactifies
the direction and turns these strings into winding strings. At the end
of this process one finds a set of modes that assemble into the
world-volume of a brane of one higher dimension, complete with the
extra component of the gauge field. This is the statement of T-duality
for multiple D-branes.

For periodic M2-branes, these will all be linked by an infinite
dimensional 3-algebra (though one does not expect membrane winding
modes when there is a single compact direction). Speculations about
the structure of the infinite 3-algebra exist in the literature (see
the comments in Ref.\;\cite{Bagger:2007vi}, following earlier ideas of
Refs.\;\cite{Basu:2004ed,Berman:2005re,Berman:2006eu} and it seems
likely that it will be simpler than a generic finite-dimensional
3-algebra. In the absence of a precise result on this, one
interpretation of our result could be that the net effect of
compactification is captured by the scalar VEV. In this interpretation
the parameter $R$ describing the VEV would be identified with the
compactification radius. Then the resulting theory should describe
D2-branes. The fact that we get $\SU(2)\times \U(1)$ for the ${\cal
A}_4$ 3-algebra would indicate that on compactification it describes
{\it two} D2-branes {\it including their centre-of-mass degree of
freedom}. This in turn would mean that before compactification it
described two M2-branes including their centre-of-mass mode, which
lies within the 3-algebra and is {\it not central} (in the sense that
it does not satisfy $[T^\phi,T^I,T^J]=0$ for all $I,J$). This picture
avoids the need to postulate triviality of the IR fixed point for two
D2-branes. Moreover it generalises in a straightforward way to
higher-rank 3-algebras (assuming they exist) and leads to an
$\SU(N)\times \U(1)$ theory, which would describe $N$ D2-branes
including their centre-of-mass mode.

An objection to this approach is that compactification of a
circle is a change of the background and should lead to a {\it
different} world-volume theory instead of the same theory with a
different VEV. However the equations we find are very suggestive that
the VEV $R$ is related to a compactification radius and we are therefore
led to suspect that our Higgsed theory in some way captures the
dynamics that would result upon compactification.

One might worry that the non-decoupling of the centre-of-mass from the
other degrees of freedom violates physical expectations following from
translation invariance. However, while the zero mode is coupled in the
3-algebra, it is not clear that this causes it to couple to {\it
physical, gauge-invariant} degrees of freedom. As an example, in the
${\cal A}_4$ 3-algebra, the scalar $X^{\phi(8)}$ that develops a VEV
breaks $\SU(2)\times \SU(2)$ to a diagonal $\SU(2)$ under which it is
neutral. Hence, as we have seen, after compactification the zero mode
does decouple from the remaining modes. But even without
compactification of a transverse direction, one would expect that the
physical degrees of freedom of the two M2-branes were contained
within this diagonal $\SU(2)$, and therefore decoupled -- but not
manifestly so -- from the putative centre-of-mass direction. For the
general case, the 3-algebra has an $\SO(N^2)$ structure in which sits
an $\SU(N)\times \SU(N)$, further broken to diagonal $\SU(N)$ when there
is a scalar VEV. The physical degrees of freedom of $N$ M2-branes,
expected to be ${\cal O}(N^{3/2})$ in number, should again sit inside
this diagonal $\SU(N)$. The overall picture is that the 3-algebra
contains vastly {\it more} ``gauge'' degrees of freedom than physical
ones, so it is quite possible for any chosen zero-mode direction to
decouple from all other modes within the physical subspace. If the
above picture is correct then decoupling will be visible only
quantum-mechanically in the correct gauge-fixed path integral, in
sharp contrast to D-branes where it is already manifest in the
classical action.\footnote{We are grateful to Shiraz Minwalla for a
discussion on this point.}

An alternative interpretation of our results is that giving a VEV to a
scalar field takes us onto a Coulomb branch where one M2-brane has
moved far away from the others, in a theory with no compactification
involved. In this case emergence of an $\SU(2)\times \U(1)$ from the
3-algebra ${\cal A}_4$ would be interpreted as describing a pair of
strongly coupled M2-branes and a decoupled M2-brane far away (as
$R\to\infty$). With this interpretation the 3-algebra ${\cal A}_4$
describes {\it three} M2-branes, as originally envisioned in
Ref.\;\cite{Bagger:2007jr}. However in contrast to
Ref.\;\cite{Bagger:2007jr} we are no longer forced to assume that the
IR fixed point on 2 M2-branes is trivial. Instead, it merely has no
3-algebra description. The situation for higher 3-algebras also has
puzzling features. In our limit, the Higgsed system is a strongly
coupled $\SU(N)$ dynamics plus a decoupled $\U(1)$ theory and this
would be interpreted as the theory on $N$ M2-branes plus a decoupled
brane. But it is not clear why the dynamics on the $N$ M2-branes is
visible as a Yang-Mills, rather than 3-algebra, theory. And the
biggest puzzle is why one needs to decouple a single M2-brane by
moving it away, in order to see the $\SU(N)$ dynamics on the
others. We expect further research will clarify the interpretation of
our result and bring about a clearer understanding of the overall
picture.

An interesting application of our methods would be to the M2-brane
theory when {\it two} transverse directions are compactified. In this
case there will be membrane winding modes somehow linking the original
membranes. This may provide a new test, as well as a better
understanding, of the 3-algebra structure. This situation is more
closely analogous to the one considered in string theory where we have
periodic arrays of D-branes connected with winding
strings\;\cite{Taylor:1996ik}. Another application would be to use the
M2 action and our D2 reduction to directly relate the Basu-Harvey
solution for the M2$\,\perp\,$M5 intersection\;\cite{Basu:2004ed} to
the D1$\,\perp\,$D3 intersection as viewed by the D1 world-volume
theory\;\cite{Constable:1999ac}. This would require compactifying in
one of the directions parallel to the M2's, itself an interesting
issue to contend with.

There have been previous attempts to carry out the reduction from
multiple M2 to D2 branes and Yang-Mills
theory\;\cite{Gustavsson:2007vu,Gustavsson:2008dy}. These papers
contain some hints of the fact that some `special' 3-algebra direction
plays a role in obtaining a Yang-Mills theory. Though the setting for
our work is a conservative one where loop algebras and non-Abelian
duality are not invoked, it would be reasonable to think of our result
as a concrete realisation of some of the nice ideas in the above
works.

Classifying 3-algebras, or at least finding a large class of them,
seems a tractable and worthwhile problem. The connection to $\SU(N)$
Lie algebras should be a useful guide. As noted above, the general
3-algebra seems to be governed by a structure like $\SO(N^2)$ with
$\SU(N)\times \SU(N)$ sitting inside it.  For $N=4$ this fits together
neatly, as locally $\SO(4)\simeq \SU(2)\times
\SU(2)$.  However, for higher $N$ it has been observed in
Ref.\;\cite{Bagger:2007vi} that there are no invariant 4th rank tensors
and therefore $f^{ABCD}$ must have a nontrivial kernel. Here we have
not addressed the question of how to find this kernel, in other words
what substructure of $\SO(N^2)$ is relevant. This is again tied to the
problem of characterising the physical, gauge invariant subspace of
the theory.

\section*{Acknowledgements}

We would like to thank James Bedford, Neil Copland, Chris Hull and
Neil Lambert for helpful discussions, and particularly Shiraz Minwalla
for several discussions and comments on the manuscript. We also wish
to thank Ashoke Sen and David Tong for their useful comments on a
previous version of this manuscript. SM is grateful for the
hospitality of the Institute for Mathematical Sciences, Imperial
College, London where part of this work was done.  The generous
support of the people of India is gratefully acknowledged.

\bibliographystyle{JHEP}
\bibliography{m2d2}

\end{document}